\documentclass[a4paper,11pt]{article}
\pdfoutput=1 

\usepackage{jcappub} 

\usepackage[T1]{fontenc} 

\title{\boldmath Strange stars in energy-momentum-conserved $f(R,T)$ gravity}

\author[a,b,1]{G.A. Carvalho,\note{Corresponding author.}}
\author[c]{S.I. dos Santos, Jr.,}
\author[a,2]{P.H.R.S. Moraes}
\author[a,2]{and M. Malheiro}

\affiliation[a]{Instituto de Pesquisa e Desenvolvimento (IP\&D), Universidade do Vale do Para\'iba, \\ 12244-000, S\~ao Jos\'e dos Campos, SP, Brazil}
\affiliation[b]{Departamento de F\'isica, Instituto Tecnol\'ogico de Aeron\'autica,\\ S\~ao Jos\'e dos Campos, SP, 12228-900, Brazil}
\affiliation[c]{Dipartamento di Fisica, Universit\`a degli Studi di Napoli Federico II, \\ Napoli 80126, Italy}

\emailAdd{geanderson.araujo.carvalho@gmail.com}
\emailAdd{samuel.isidoro.santos@gmail.com}
\emailAdd{moraes.phrs@gmail.com}
\emailAdd{malheiro@ita.br}

\abstract{For the accurate understanding of compact objects such as neutron stars and strange stars, the Tolmann-Openheimer-Volkof (TOV) equation has proved to be of great use. Hence, in this work, we obtain the TOV equation for the energy-momentum-conserved $f(R,T)$ theory of gravity to study strange quark stars. The $f(R,T)$ theory is important, especially in cosmology, because it solves certain incompleteness of the standard model. In general, there is no intrinsic conservation of the energy-momentum tensor in the $f(R,T)$ gravity. Since this conservation is important in the astrophysical context, we impose the condition $\nabla T_{\mu\nu}=0$, so that we obtain a function $f(R,T)$ that implies conservation. This choice of a function $f(R,T)$ that conserves the momentum-energy tensor gives rise to a strong link between gravity and the microphysics of the compact object. We obtain the TOV by taking into account a linear equation of state to describe the matter inside strange stars, such as $p=\omega\rho$ and the MIT bag model $p=\omega(\rho-4B)$. With these assumptions it was possible to derive macroscopic properties of these objects.}

\begin{document}
\maketitle
\flushbottom

\section{Introduction} \label{intro}

Alternative theories of gravity have the purpose of solving some issues that, in principle, General Relativity cannot, such as the dark energy  \cite{weinberg/1989,hinshaw/2013} and dark matter problems \cite{danieli/2019} and even the theoretical prediction of observed massive pulsars \cite{demorest/2010,antoniadis/2013} and super-Chrandrasekhar white dwarfs \cite{howell/2006,scalzo/2010,kepler/2007} that hardly can be explained assuming standard structure and equation of state (EoS) for these objects. 

Today, probably the most popular of the alternative gravity theories is the $f(R)$ theory \cite{nojiri/2011,sotiriou/2010,de_felice/2010}, which takes a general function of the Ricci scalar $R$ in the gravitational action as its starting point. Indeed, the presence of general terms in $R$ in the action yields extra terms in the field equations of the theory, and those, in a cosmological aspect, can explain the present cosmic acceleration \cite{riess/1998,perlmutter/1999} with no need for dark energy \cite{amendola/2007,song/2007}. Such extra terms can also elevate the maximum mass theoretically expected for neutron stars \cite{capozziello/2016,astashenok/2013,astashenok/2015} and white dwarfs \cite{das/2015}. Anyhow, some $f(R)$ gravity flaws in the solar system scale were reported, for instance, in \cite{erickcek/2006,chiba/2007,capozziello/2007} and should discard most of the $f(R)$ models proposed so far. For the galactic scales, the $f(R)$ theory also does not seem to be suitable  \cite{dolgov/2003,chiba/2003,olmo/2005}.

In \cite{harko/2011}, it was proposed a generalization of the $f(R)$ theory, by including in the gravitational action, besides the general term in $R$, a general term in $T$, the trace of the energy-momentum tensor, yielding the $f(R,T)$ gravity. The $T$-dependence on such a theory may be due to the existence of imperfect fluids in the universe and could generate a theory that involves gravity and quantum mechanics \cite{xu/2016}. 

The $f(R,T)$ theory describes pretty well the solar system regime \cite{shabani/2014}. New terms coming from this theory attend to describe dark matter galactic effects \cite{zaregonbadi/2016}. It was also shown that $f(R,T)$ gravity can give a considerable contribution to the gravitational lensing \cite{alhamzawi/2016} and a deviation to the usual geodesic equation \cite{baffou/2017}.

Moreover, the $f(R,T)$ cosmology evades the dark energy problem, by describing the cosmic acceleration as due to the extra terms in $T$ in the field equations of the model \cite{mrc/2016,ms/2017}, instead of being due to the presence of the cosmological constant.

In opposition to other alternative gravity theories, the modifications here in this theory are associated to material terms instead of geometric ones. This new terms yield the non-vanishing of the covariant derivative of the matter energy-momentum tensor, that is,  $\nabla_\mu T^{\mu\nu}\neq0$ \cite{harko/2011,barrientos/2014}. 
The fact that the energy-momentum tensor is not  conserved in this theory can be related, in a cosmological perspective, to creation (or destruction) of matter during the universe evolution. This subject was investigated from a thermodynamical perspective \cite{harko/2014}. The same kind of physical property can be noticed in other non-conservative energy-momentum theories, such as those presented in \cite{harko/2010,harko/2013}.

In astrophysics, particularly in the study of hydrostatic equilibrium configurations of stellar objects, the association of the non-vanishing of the energy-momentum tensor covariant derivative with matter creation is not correct since the Tolman-Oppenheimer-Volkoff (TOV) equation \cite{tolman/1939,oppenheimer/1939} is worked out in a static regime and we do not know exactly what would be the right mechanism to create matter inside a star.

Therefore, instead of searching for a physical interpretation for this issue, one could attempt to construct a TOV equation from a ``conservative'' version of the $f(R,T)$ gravity. In fact, the energy-momentum conserved (EMC) version of $f(R,T)$ gravity has been worked out in the literature within different approaches, such as neutron stars hydrostatic equilibrium and even cosmology, and we are going to visit those later (check Section \ref{sec:2}).

Our purpose here is to take one step further within the EMC $f(R,T)$ gravity, by constructing the hydrostatic configurations of strange stars (SSs) \cite{itoh/1970,alcock/1986,haensel/1986} in the formalism. As we are going to revisit, the function $h(T)$ that conserves the energy-momentum tensor within $f(R,T)=R+h(T)$  depends on the EoS of matter, which in the present case we assume to be the EoS of strange quark matter.

We naturally wish to confront our results with some observational data of SSs \cite{li/1999,glendenning/1989,li/2011,bombaci/2002}. We will also compare them with other SS models constructed in alternative gravity, such as those obtained from $f(\mathcal{T})$ gravity \cite{abbas/2015}, with $\mathcal{T}$ being the torsion scalar, $f(R)$ gravity \cite{panotopoulos/2017} and even the non-conservative case of $f(R,T)$ gravity \cite{deb/2018,biswas/2019,debabrata/2019}.

This article is organized as follows: in Section \ref{sec:1}, we describe some important mathematical and physical properties of the $f(R,T)$ gravity. In Section \ref{sec:2} we discuss about the EMC $f(R,T)$ gravity already present in the literature. We derive the EMC $f(R,T)$ gravity for SSs in Section \ref{sec:sec3}. We present and solve the referred TOV-like equations in Section \ref{sec:tov}. We highlight and discuss our results in Section \ref{sec:con}.

\section{The $f(R,T)$ gravity}
\label{sec:1}

In order to obtain the field equations of the $f(R,T)$ gravity theory, one starts from the following action \cite{harko/2011}

\begin{equation}\label{frt1}
\mathcal{S}=\int\left[\frac{f(R,T)}{16\pi}+\mathcal{L}_m\right]\sqrt{-g}d^{4}x,
\end{equation}
with $f(R,T)$ being a general function of $R$ and $T$, $\mathcal{L}_m$ being the matter lagrangian density, $g$ the determinant of the metric $g_{\mu\nu}$ and natural units are assumed throughout the paper.

When varying such an action with respect to $g_{\mu\nu}$ we obtain:

\begin{equation}\label{frt2}
G_{\mu\nu}=8\pi T_{\mu\nu}+\frac{1}{2}h(T)g_{\mu\nu}+h_T(T)(T_{\mu\nu}-\mathcal{L}_mg_{\mu\nu}),
\end{equation}
in which $G_{\mu\nu}$ is the Einstein tensor, $T_{\mu\nu}$ is the energy-momentum tensor and we have considered $f(R,T)=R+h(T)$, with $h(T)$ being a function of $T$ only, so that one recovers General Relativity in the regime $h(T)=0$. Moreover, $h_T(T)\equiv dh(T)/dT$.

The covariant derivative of $T_{\mu\nu}$ in (\ref{frt2}) is

\begin{equation}\label{frt3}
\nabla^{\mu}T_{\mu\nu}=\frac{h_T(T)}{8\pi+h_T(T)}\left[(\mathcal{L}_mg_{\mu\nu}-T_{\mu\nu})\nabla^{\mu}\ln h_T(T)+\nabla^{\mu}\left(\mathcal{L}_m-\frac{1}{2}T\right)g_{\mu\nu}\right].    
\end{equation}
From \eqref{frt3} we can see the previously mentioned non-conservation of the energy-momentum tensor in $f(R,T)$ gravity. In the next section we are going to briefly review some applications of EMC $f(R,T)$ gravity.

\subsection{The energy-momentum conserved formalisms proposed for the $f(R,T)$ gravity}\label{sec:2}

Some different approaches have already been made searching for EMC cases of the $f(R,T)$ theory. Looking for Eq.\eqref{frt3}, we see that there are at least two possibilities to turn $f(R,T)$ gravity into an EMC theory. Let us briefly present these possibilities and their applications below.

In \cite{mcr/2018} it was proposed a form to conserve the energy-momentum tensor in $f(R,T)=R+2\lambda T$ cosmology, with $\lambda$ a constant. In order to illustrate that, the $f(R,T)=R+2\lambda T$ field equations, obtained from the substitution of $h(T)=2\lambda T$ in \eqref{frt2}, were rewritten as 

\begin{equation}\label{ssf1}
G_{\mu\nu}=8\pi T_{\mu\nu}^{\texttt{eff}},
\end{equation}
with $T_{\mu\nu}^{\texttt{eff}}=T_{\mu\nu}+\tilde{T}_{\mu\nu}$ and 

\begin{equation}\label{ssf2}
\tilde{T}_{\mu\nu}\equiv\frac{\lambda}{8\pi}[2(T_{\mu\nu}-\mathcal{L}_mg_{\mu\nu})+Tg_{\mu\nu}].    
\end{equation}

Within such a formalism, the application of the Bianchi identities in \eqref{ssf1} yields $\nabla^\mu[8\pi(T_{\mu\nu}+\tilde{T}_{\mu\nu})]=0$ or simultaneously $\nabla^\mu T_{\mu\nu}=0$ and $\nabla^\mu\tilde{T}_{\mu\nu}=0$. The first of these two equations yields the usual conservation law of standard cosmology while the second yields the EoS of stiff matter \cite{chavanis/2015}. In this way, the $f(R,T)$ gravity indicated the existence of a two-fluid cosmological model, in which each of the fluids is conserved during the universe evolution.

S. Chakraborty, on the other hand, has shown that a part of the arbitrary function $f(R,T)$ can be determined if one imposes $\nabla_\mu T^{\mu\nu}=0$ \cite{chakraborty/2013}. A cosmological model was derived from such a principle \cite{alvarenga/2013}. In \cite{alvarenga/2013}, in order to obtain the EMC cosmological model, the authors have assumed a Friedmann-Lema\^itre-Robertson-Walker metric as well as the EoS $p=\omega\rho$, with $p$ being the pressure, $\omega$ the constant EoS parameter and $\rho$ the density of the universe. By solving \eqref{frt3}, then, they have found $h(T)\sim T^\frac{1+3\omega}{2(1+\omega)}$.

A similar approach was recently applied to the TOV equation for neutron stars \cite{scmm/2018}, that is, a conservative function $h(T)$ was found for the polytropic EoS \cite{tooper/1964} case and the TOV-like equation was constructed and solved from such a formalism. It is important to remark that the conservative case has better results in comparison with the non-conservative version of the TOV equation within $f(R,T)$ gravity \cite{mam/2016}. While the contribution of the $f(R,T)$ gravity for neutron stars in the latter case resulted in slightly greater masses and greater radii, in the former conservative case, the maximum masses of neutron stars were substantially increased ($>2M_\odot$) while their radii did not vary significantly, getting in touch with massive pulsars observations \cite{demorest/2010,antoniadis/2013}.

Another advantage of the EMC $f(R,T)$ gravity can be seen in the realm of cosmology, in which such a model is clearly in advantage when compared to the non-conservative cases for what concerns the confrontation of theoretical predictions with supernovae Ia observational data \cite{velten/2017}.

\section{Energy-momentum-conserved $f(R,T)$ gravity for strange quark matter}\label{sec:sec3}

Now we wish to construct an EMC model for $f(R,T)$ gravity to be used to obtain the hydrostatic equilibrium configurations of SSs. SSs are stars that contain superdense matter on its fundamental level \cite{itoh/1970,alcock/1986,haensel/1986}, that is, strange matter. We still do not know if this is, indeed, the fundamental level of matter at high densities and that is exactly what makes the study of SSs so important. Anyhow, some SSs candidates are well known \cite{li/1999,glendenning/1989,li/2011,bombaci/2002} as well as some methods to prove SSs and, consequently, strange matter existence via gravitational wave astronomy \cite{andersson/2002,geng/2015,mm/2014,zhou/2018,malheiro4/2007}.

In order to start the construction of the EMC $f(R,T)$ gravity for SSs, let us work with Eq.\eqref{frt3} by forcing $\nabla^\mu T_{\mu\nu}=0$ on it. By assuming $\mathcal{L}_m=\rho$ and $\mu=1$ yields the following differential equation:
\begin{equation}\label{ssf3}
(\rho+p)({\rm ln}h_T)' +\frac{1}{2}(\rho+3p)'=0,
\end{equation}
where the comma stands for radial derivative.

By assuming the EoS to be $p=\omega\rho$ to describe the matter inside such objects and solving Eq.\eqref{ssf3} yields

\begin{equation}\label{ssf4}
h_T=\frac{1}{2}\left(\frac{1-\omega}{1+\omega}\right)\lambda T^{-\frac{1}{2}\left(\frac{1+3\omega}{1+\omega}\right)},
\end{equation}
with $\lambda$ being an arbitrary constant.

By integrating \eqref{ssf5}, one has

\begin{equation}\label{ssf5}
h(T)=\lambda T^{\frac{1}{2}\left(\frac{1-\omega}{1+\omega}\right)}.
\end{equation}
We observe from \eqref{ssf5} that for $\omega=0$, $h(T)\sim\sqrt{T}$, which is the same result obtained for an EMC $f(R,T)$ gravity cosmological model in the case of pressureless ($\omega=0$) matter \cite{alvarenga/2013}. Moreover, since $0<\omega<1$, the exponent of $T$ in \eqref{ssf5} is restricted to values between 0 and 1/2.

Let us now call the MIT bag model EoS \cite{alcock/1986,haensel/1986,malheiro3/2016,malheiro0/2015,debabrata/2019,malheiro2/2003,negreiros/2009} to describe matter inside SSs in the EMC $f(R,T)$ gravity model. Such an EoS describes a fluid composed of up, down and strange quarks only. The relation between pressure and energy density becomes a linear one, given by $p=\omega(\rho-4B)$, with constant $\omega$ and $B$ being the bag constant.

By using the MIT bag model EoS, the EMC functional form $h(T)$ is calculated from the integration of

\begin{equation}\label{functder}
  h_T(T)=\beta\frac{(1+\omega)}{(1-3\omega)} \left[(T-12B\omega)\frac{(1+\omega)}{(1-3\omega)}-4B\omega\right]^{-\frac{1}{2}\frac{(1+3\omega)}{(1+\omega)}}
\end{equation}
and reads

\begin{equation}\label{funct}
  h(T)=\beta\left[(T-12B\omega)\frac{(1+\omega)}{(1-3\omega)}-4B\omega\right]^{\frac{(1-\omega)}{2(1+\omega)}},
\end{equation}
where $\beta$ is an arbitrary constant.
\section{The Tolman-Oppenheimer-Volkoff equations and their solutions}\label{sec:tov}

Let us now use Eq.\eqref{ssf5} to construct the TOV-like equation in this model. By using the spherical static metric, 

\begin{equation}\label{metric}
ds^2=e^{\phi} dt^2-e^{\psi}dr^2-r^2d\theta^2-r^2\sin^2\theta d\phi^2,
\end{equation}
with $\phi=\phi(r)$ and $\psi=\psi(r)$ being the metric potentials, as well as assuming the energy-momentum tensor of a perfect fluid, we obtain the $00$ and $11$ components of the field equations as

\begin{subequations}\label{eq:camp:3}
	\begin{align}
		\begin{split}
			&\frac{e^{-\psi}}{r^{2}}(e^{\psi}+\psi'r-1)=8\pi \rho + \frac{1}{2}h,
		\end{split}
		\label{eq:camp:tt}
		\\
		\begin{split}
			\frac{e^{-\psi}}{r^2}\left(1-e^{\psi}+\phi'r\right)=8\pi p-\frac{1}{2}h-h_T(p-\rho).
		\end{split}
		\label{eq:camp:rr}
	\end{align}
\end{subequations}

We introduce now the quantity $m=m(r)$, such that

\begin{equation}
	e^{-\psi}=1-\frac{2m}{r},
\end{equation}
and replacing it into \eqref{eq:camp:tt}, we get

\begin{equation}\label{masseq}
	\frac{dm}{dr}=4\pi \rho r^2 + \frac{1}{4}hr^2,
\end{equation}
so that $m(r)$ represents the enclosed gravitational mass within a sphere of radius $r$ according to the EMC $f(R,T)$ gravity.

Let us recall that from the conservation of the energy-momentum tensor we have:

\begin{equation}
	\nabla^{\mu}T_{\mu\nu}=-p' -(\rho+p)\frac{\phi'}{2}=0.
\end{equation}

By isolating $\phi'$ in \eqref{eq:camp:rr} and substituting in the above equation, one is able to derive the modified TOV equation as follows

\begin{equation}\label{tovlike}
	p'=-(\rho+p)\dfrac{\left\{\frac{m}{r^2}+ \left[4\pi p - \frac{1}{4}h -\frac{1}{2}h_{T}(p-\rho)\right]r\right\}}{\left(1-\frac{2m}{r}\right)}.
\end{equation}

The Equations \eqref{masseq} and \eqref{tovlike} can be solved numerically by using the fourth-order Runge-Kutta method and considering a specific model for the functional $h$ and $h_T$. In order to do so, the boundary conditions at the center of the star are as follows: $p(0)=p_c$, $\rho(0)=\rho_c$ and $m(0)=0$, with $p_c$ and $\rho_c$ being the central pressure and central energy density. For $r=R$, where the pressure and energy density of the star vanish, the enclosed mass $m(R)=M$ represents the total mass of the star, with $R$ being its total radius. By using different values of central energy density, one is able to construct the mass-radius relation as well as other relations that we further derive in this work.

\subsection{Case: $p=\omega \rho$}

Let us consider now the first case derived in Section \ref{sec:sec3}, where an EoS like $p=\omega\rho$ was used to obtain the functionals $h$ and $h_T$. For this case the TOV-like equation becomes

\begin{equation}\label{newtov}
	p'=-(\rho+p)\frac{\left\{\frac{m}{r^2}+ \left[4\pi p +\zeta(\omega)\rho^{\frac{1-\omega}{2(1+\omega)}}\right]r\right\}}{\left(1-\frac{2m}{r}\right)},
\end{equation}
where the parameter $\zeta(\omega)$ is given by

\begin{equation}\label{zeta}
	\zeta(\omega)=\frac{\lambda}{2}\left[(1+8\pi)(1-3\omega)^{-\frac{1+3\omega}{2(1+\omega)}}-\frac{1}{2} (1-3\omega)^{\frac{1-\omega}{2(1+\omega)}} \right],
\end{equation}
such that $\lambda=0$ yields the usual TOV equation. It is worth to note that $\omega=\frac{1}{3}$ also gives the standard TOV equation, which is expected since this value for $\omega$ yields $T=0$, hence cancelling out any contribution from the trace of the energy-momentum tensor in the field equations.

Figure \ref{massxradius} below shows the behaviour of the total mass with total radius of the star, where five values of $\lambda$ were used and $\omega=0.28$ in reference to the MIT bag model EoS  with $B=0$. For this case of $\omega$, the approach presented in Section \ref{sec:2} yields $h(T)=\lambda T^{0.28}$. It is worth to quote that $\lambda=0$ corresponds to the result found within General Relativity framework. 

\begin{figure}[h!]
	\begin{center}
		\includegraphics[width=0.6\linewidth]{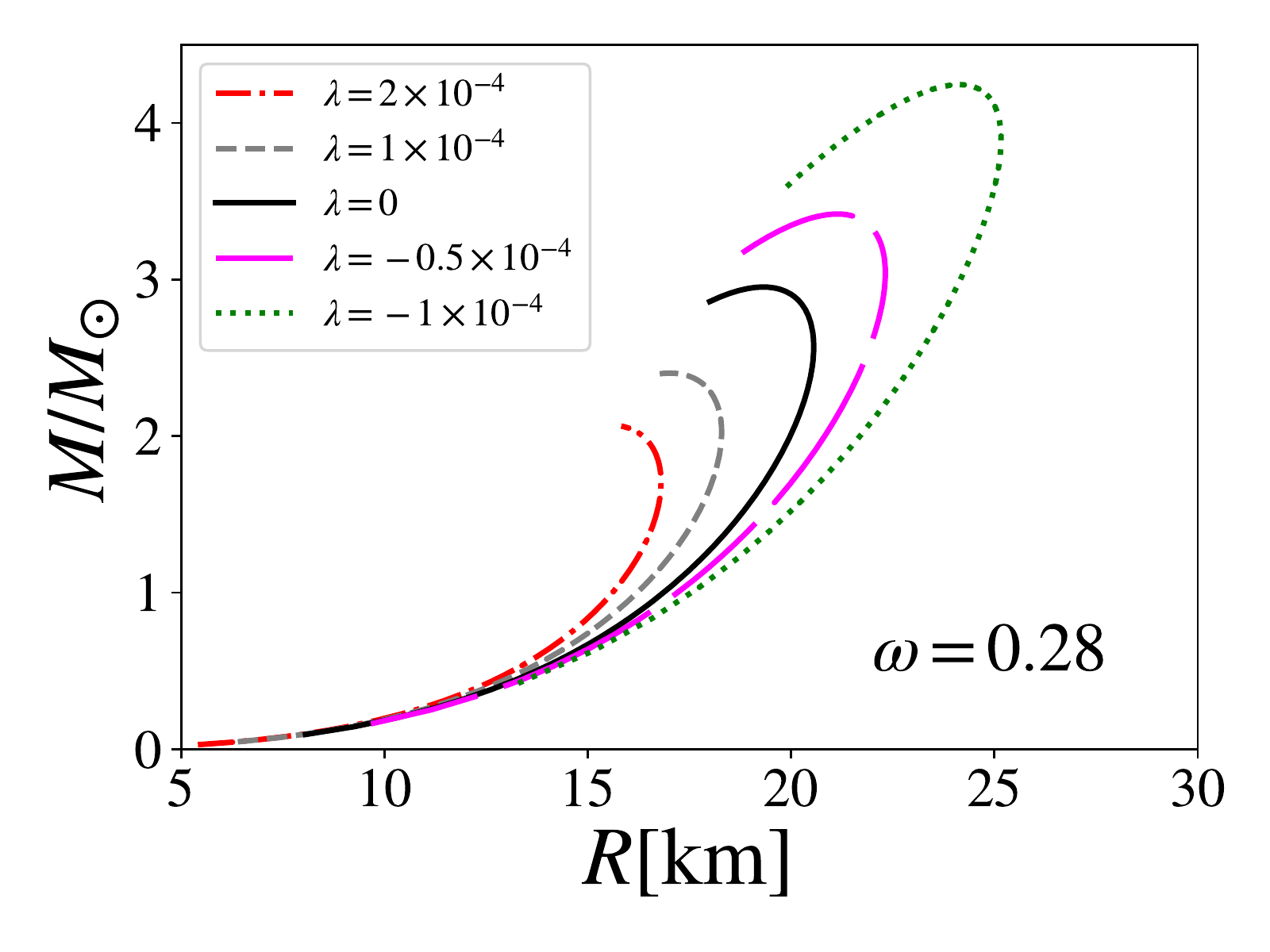}
		\caption{Mass-radius relation for $p=\omega\rho$ within the conservative model of the $f(R,T)$ gravity for the interval $50-800$ MeV/fm$^{3}$ for the central density. Several values of $\lambda$ were employed and $\omega=0.28$.}
		\label{massxradius}
	\end{center}
\end{figure}

From Figure \ref{massxradius} we can note that for $\lambda>0$, less massive and smaller stars are found. Such a behaviour can be understood as a strong gravity regime effect, such like the case of General Relativity Theory in comparison with Newtonian gravitation \cite{carvalho/2018}.  On the other hand, for the cases where $\lambda<0$ we observe an increasing in the total mass and total radius of the star when $\vert\lambda\vert$ increases. We can see that the increasing in the mass and radius of the star is very sensitive to the value of $\vert\lambda\vert$. 

\begin{figure}
\includegraphics[width=0.49\linewidth]{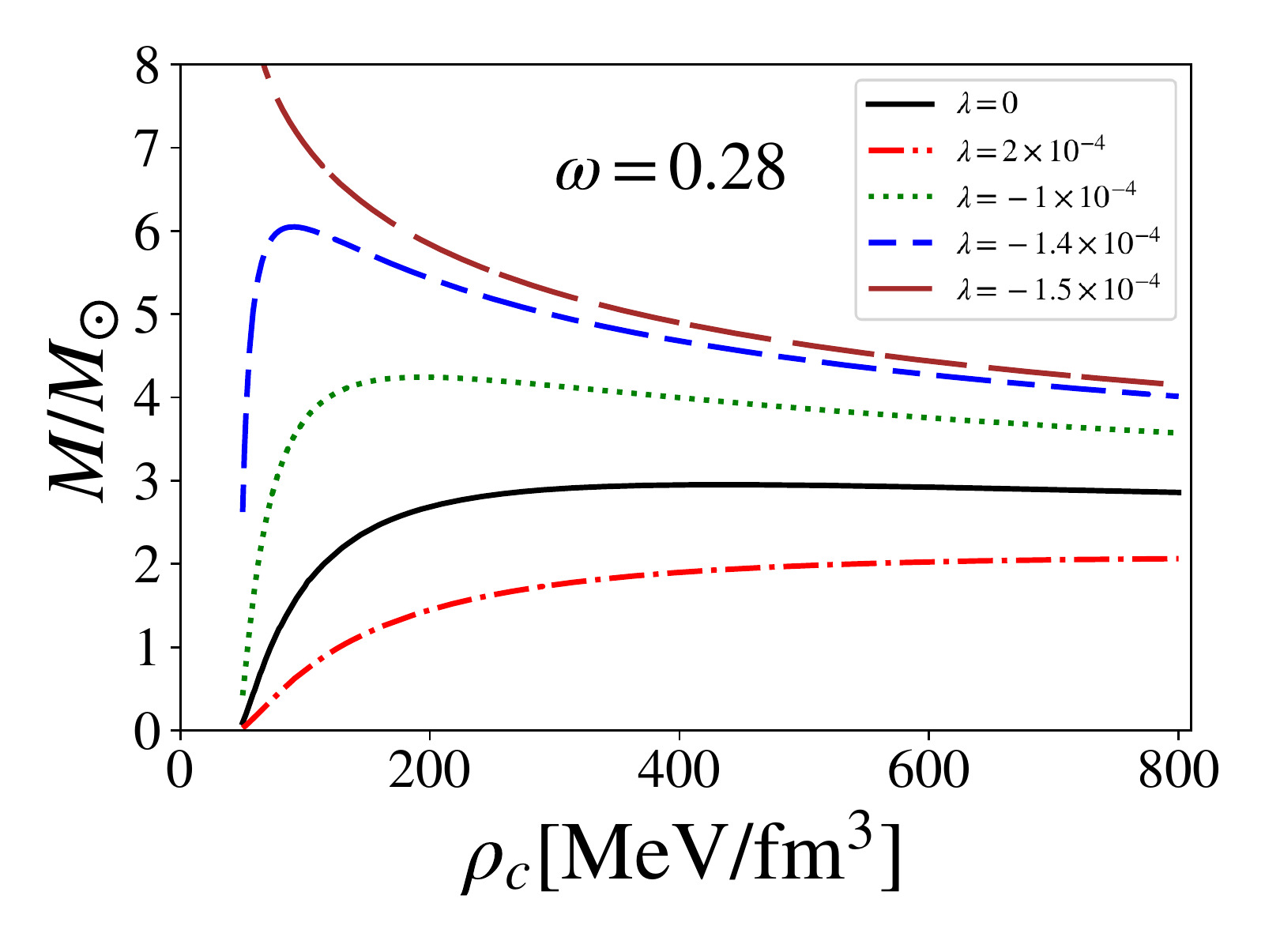}
\includegraphics[width=0.49\linewidth]{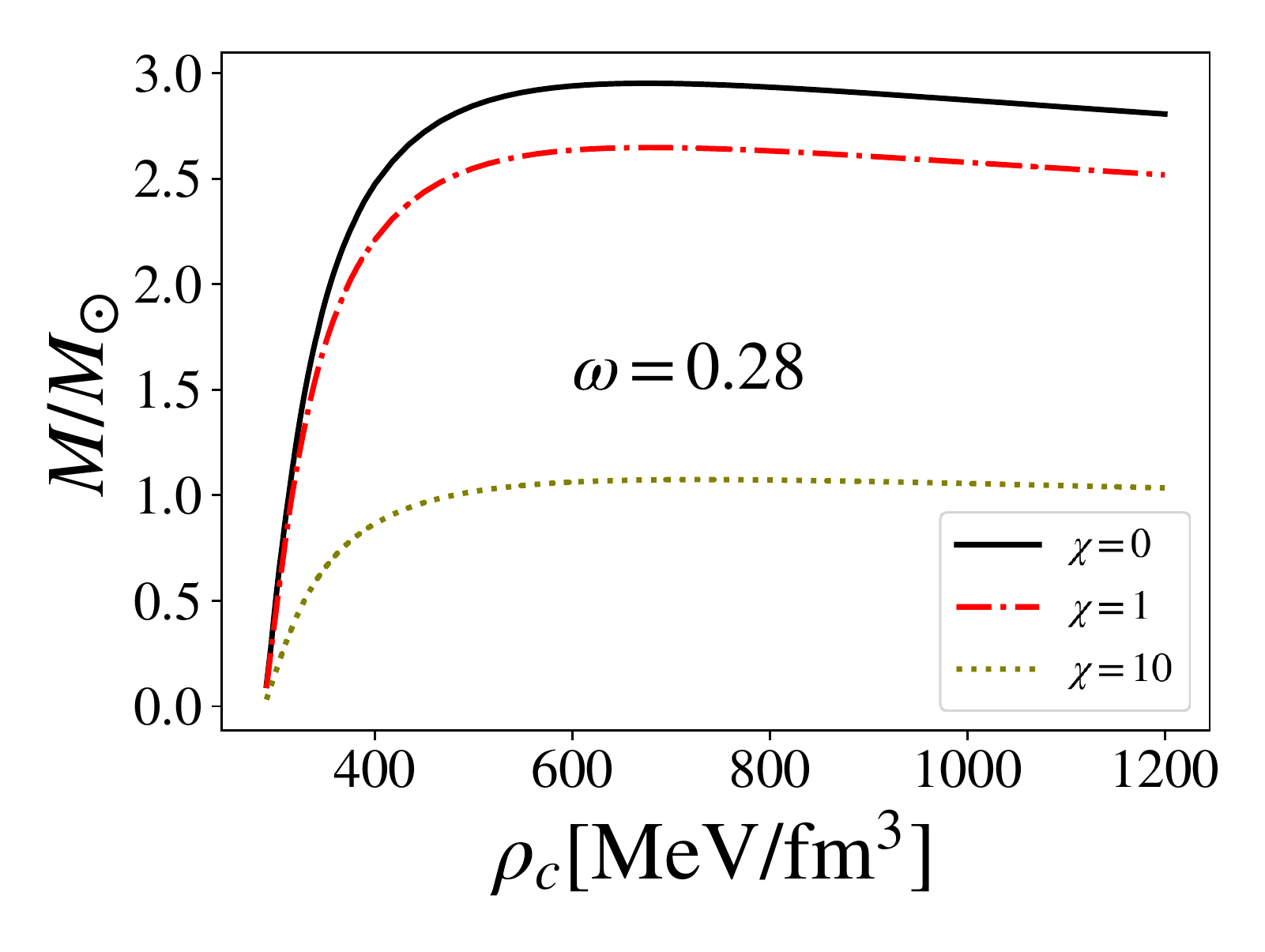}
\caption{Left panel: total mass versus central energy density for the conservative model of the $f(R,T)$ gravity using several values of $\lambda$, $\omega=0.28$ and $B=0$. The value $\lambda=-1.5\times 10^{-4}$, in the interval of central energy density we have used ($50-800~\rm Mev/fm^3$), does not produce any stable stars. Right panel: total mass versus central density for the non-conservative linear model of $f(R,T)$ gravity, i.e., $f(R,T)=R+2\chi T$. Some values of $\chi$ and $\omega=0.28$ were used.}
	\label{massxdensity2}
\end{figure}

In the left panel of Fig.\ref{massxdensity2} below we show the mass against central energy density for the conserved model of $f(R,T)$ gravity, where several values of $\lambda$, $\omega=0.28$ and $B=0$ were employed. For the values of $\lambda$ larger than $-1.5\times 10^{-4}$ we observe that the mass initially increases with central density until it attains a maximum value. After that point, the mass decreases with the increasing of central density. 

From the regular criterion of stability, $\partial M/\partial\rho_c>0$, we conclude that the maximum mass points mark the onset of instability in each of those curves of Fig.\ref{massxdensity2}. However, the value of $\lambda=-1.5\times 10^{-4}$ does not produce any stable stars in the considered range of central energy density $(50-800~\rm Mev/fm^3)$ since its mass-density relation does not respect the stability criterion, in this way, setting up a lower limit for $\lambda$ and a maximum stable mass of  $\sim6M_{\odot}$. 

In the right panel of Fig.\ref{massxdensity2} we consider the same EoS, with $\omega=0.28$, in the context of a linear non-conservative $f(R,T)$ gravity model, namely, $f(R,T)=R+2\chi T$ gravity, with $\chi$ a free parameter, as the one discussed in \cite{das/2016}. 

\subsection{Case: MIT Bag Model $p=\omega (\rho-4B)$}

Now, using the MIT bag model EoS we derive a different TOV-like equilibrium equation as

\begin{eqnarray}\label{allnewtov}
	\frac{dp}{dr}=-\dfrac{(\rho+p)}{\left(1-\frac{2m}{r}\right)}\left\{\frac{m}{r^2}+ \left[4\pi p - \beta\xi(\omega,\rho)(p-\rho)\right]r\right\},
\end{eqnarray}
where $\xi(\omega,\rho)$ is given by

\begin{equation}
\xi(\omega,\rho)=\frac{1}{2}\left\{\frac{1}{2}[\rho(1+\omega)-4\omega B]^{\frac{1-\omega}{2(1+\omega)}}-\frac{1+\omega}{1-3\omega}[-\rho(1+\omega)+4\omega B]^{-\frac{1}{2}\frac{1+3\omega}{1+\omega}}\right\}.
\end{equation}
We will consider $\omega=0.28$, which gives a quark mass of 250 MeV/fm$^3$, and the bag constant will be  taken as $B=60$ MeV/fm$^3$. 

\begin{figure}[h!]
\begin{center}
	\includegraphics[width=0.49\linewidth]{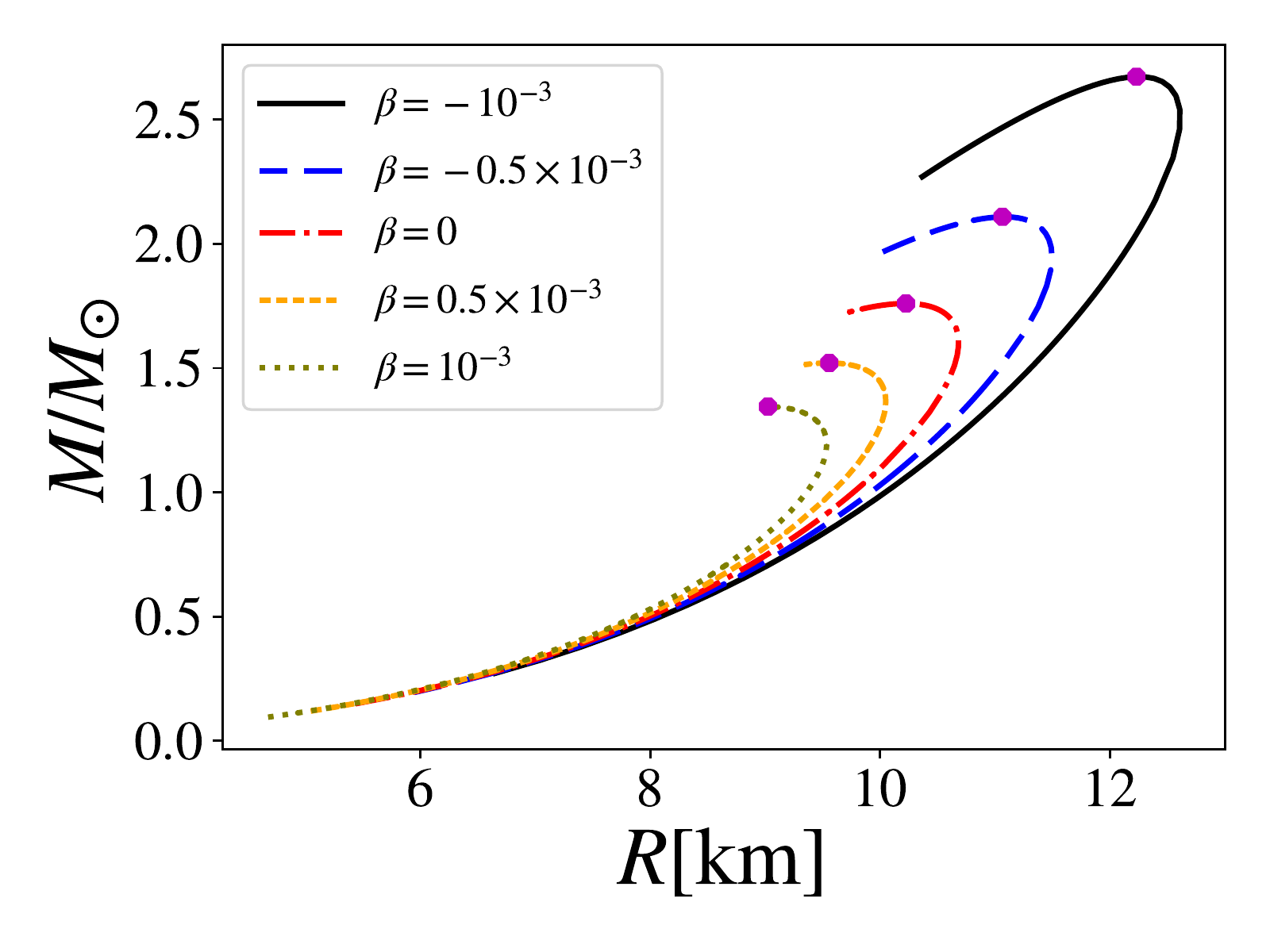}
    \includegraphics[width=0.49\linewidth]{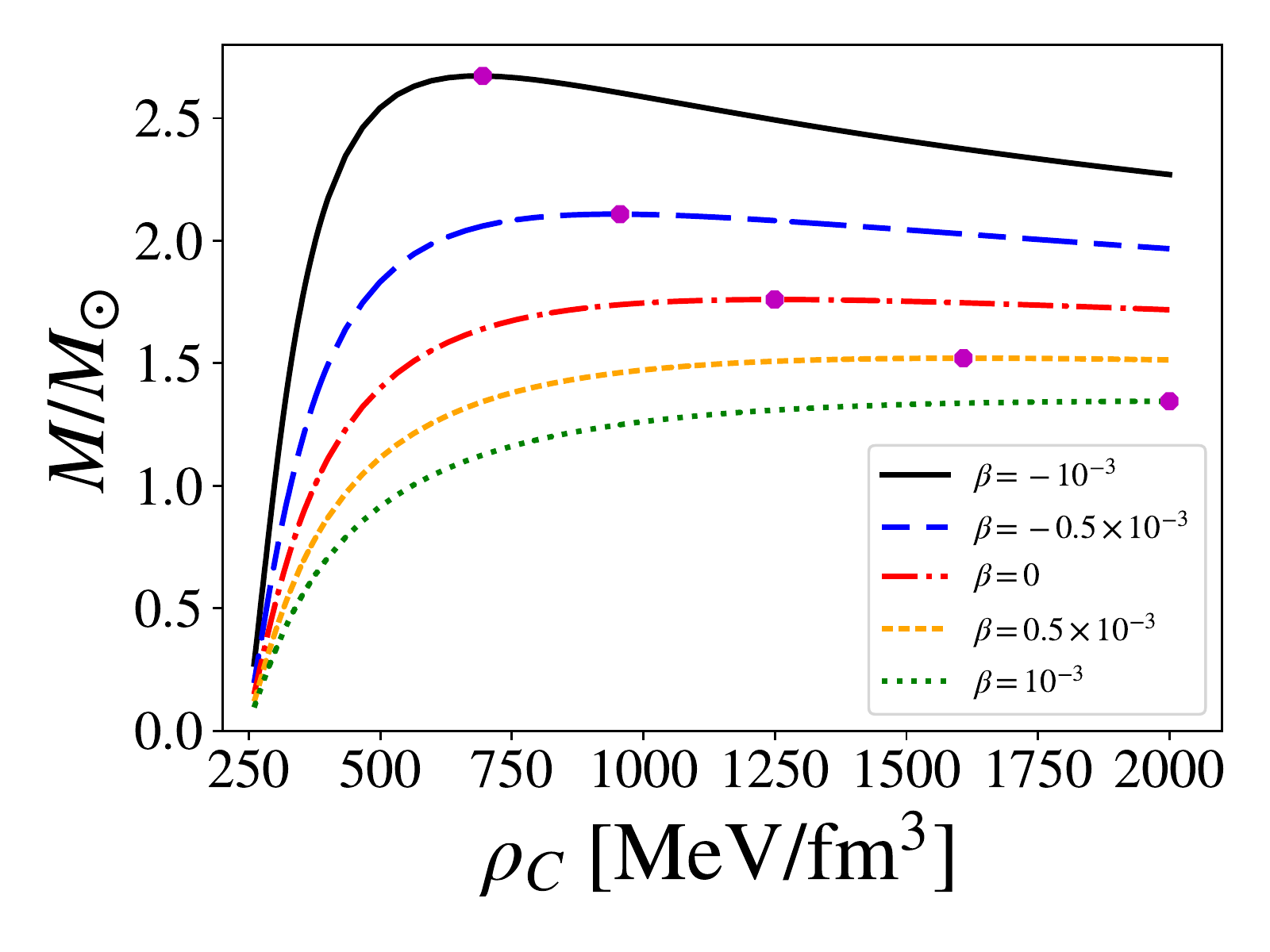}
	\caption{\label{fig:SS} Left panel: Mass-radius relation for the MIT bag model of the EMC $f(R,T)$ gravity. Right panel: Mass versus central energy density for the EMC $f(R,T)$ gravity using the MIT bag model EoS. Different values of $\beta$, $\omega=0.28$ and $B=60$ MeV/fm$^3$ were employed on both plots.}
	\label{massxradiusmit}
\end{center}
\end{figure}

The mass-radius and mass-density relations are shown in Fig.\ref{fig:SS}. The considered values for $\beta$ range from $-1\times 10^{-3}$ to $1\times 10^{-3}$, and the value $\beta=0$ corresponds to the results found within General Relativity framework. It can be seen that the positive values of $\beta$ tend to reduce the star mass and shrink the star radius, which can be understood as a gravitational force ``stronger'' than the General Relativity one (when $\beta=0$). The opposite behavior is found for negative $\beta$, where larger and more massive stars are found and this behavior can be understood as a ``weaker'' gravitational force. One interesting feature of assuming negative values of $\beta$ is that it allows a larger maximum mass. For instance, the maximum mass for $\beta=-10^{-3}$ is $M_{\rm max}\approx 2.6 M_{\odot}$. On the other hand, for the case of $\beta=0$ the maximum mass is $\sim 2 M_{\odot}$, which represents a value $\sim30\%$ smaller. 

\begin{table*}
  \begin{center}
    \caption{\label{tab1} Physical parameters of observed strange star candidates derived using $\beta=-0.5\times 10^{-3}$ and $B=60$MeV/fm$^3$. $Z_s$ represents the surface redshift and it is calculated as: $Z_s=\frac{1}{\sqrt{1-2M/R}}-1$.}
    \begin{tabular}{ccccc} 
        \hline
      SS candidate & Observed mass $M/M_{\odot}$ & Predicted radius (km) & $M/R$ & $Z_s$\\
        \hline
       PSR J1614-2230  & 1.97$\pm$0.04 \cite{demorest/2010}  & $11.49^{+0.02}_{-0.01}$ & 0.253  & 0.423\\
       Vela X-1  & 1.77$\pm$0.08 \cite{rawls/2011}  & $11.40^{+0.07}_{-0.08}$ & 0.229  & 0.359\\
       4U 1608-52  & 1.74$\pm$0.14 \cite{guver/2010}  & $11.33^{+0.06}_{-0.12}$ & 0.227  & 0.353\\
       PSR J1903+327 & 1.667$\pm$0.021 \cite{freire/2011} & $11.28^{+0.07}_{-0.04}$ & 0.218  & 0.332\\
       4U 1820-30  & 1.58$\pm$0.06 \cite{guver/2010a} & $11.18^{+0.08}_{-0.1}$ & 0.209  & 0.310\\
       Cen X-3  & 1.49$\pm$0.08 \cite{rawls/2011} & $11.03^{+0.12}_{-0.14}$ & 0.199  & 0.290\\
       EXO 1785-248  & 1.3$\pm$0.2 \cite{ozel/2009} & $10.65^{+0.38}_{-0.47}$ & 0.180 & 0.250\\
       LMC X-4  & 1.29$\pm$0.05 \cite{rawls/2011} & $10.64^{+0.12}_{-0.11}$ & 0.179  & 0.248\\
       SMC X-1  & 1.04$\pm$0.09 \cite{rawls/2011} & $10.04^{+0.22}_{-0.26}$ & 0.153 & 0.200\\
       SAX J1808.4-3658  & 0.9$\pm$0.3 \cite{elebert/2009} & $9.62^{+0.83}_{-1.13}$ & 0.138  & 0.175\\
       4U 1538-52  & 0.87$\pm$0.07 \cite{rawls/2011}  & $9.54^{+0.21}_{-0.27}$ & 0.135 & 0.169\\
       HER X-1  & 0.85$\pm$0.15 \cite{abu/2008}  & $9.49^{+0.46}_{-0.57}$ & 0.132  & 0.166\\
       \hline
    \end{tabular}
  \end{center}
\end{table*}

\begin{table*}
  \begin{center}
    \caption{\label{tab2} Physical parameters of the strange star candidate LMC X-4 for different values of $\beta$ and $B=60$MeV/fm$^3$.}
    \begin{tabular}{cccc} 
        \hline
      $\beta$ & Predicted radius (km) & $M/R$ & $Z_s$\\
        \hline
       $-10^{-3}$ & 10.84$^{+0.11}_{-0.12}$ & 0.176 & 0.242 \\    
       $-0.5\times 10^{-3}$ & $10.64^{+0.12}_{-0.11}$ & 0.179  & 0.248\\
       $0$  & 10.38$^{+0.08}_{-0.09}$ & 0.184 & 0.257 \\
       $0.5\times 10^{-3}$ & 10.02$^{+0.03}_{-0.05}$ & 0.19 & 0.270 \\
       $10^{-3}$  & 9.44$^{+0.28}_{-0.07}$ & 0.212 & 0.295 \\
       \hline
    \end{tabular}
  \end{center}
\end{table*}

Strange quark stars have been studied in the non-conserved $f(R,T)$ theories in several recent works by using the MIT bag model equation of state \cite{das/2016,debabrata/2019,deb/2018,mam/2016,deb/2018b}. In this work, rather than consider a non-conserved $f(R,T)$ theory of gravity we derived the conserved form of the theory by using also the MIT equation of state. The outcomes of our conserved model can be compared with observed parameters of strange star candidates. Table \ref{tab1} shows the observed mass of strange star candidates and the predicted radius, compactness and gravitational redshift with use of the energy-momentum conserved $f(R,T)$ gravity for the value of $\beta=-0.5\times 10^{-3}$, and table \ref{tab2} shows also the predicted radius, compactness and gravitational redshift for the star LMC X-4 for several values of $\beta$. From table \ref{tab2} one can see that the increasing of the parameter $\beta$ yields to smaller radius for the object LMC X-4, which means that the object would be more compact and with a larger surface gravitational redshift. However, when $\beta$ is negative the increasing in its magnitude leads to larger radii and hence the behavior of the compactness and redshift are reversed. Table \ref{tab1} also allows us to confirm the feasibility of our work concerning the confrontation with observational data of compact objects.    

\section{Discussion}\label{sec:con}

Recent theoretical studies show that alternative theories to General Relativity provide important insights to solve complex issues of present astrophysical and cosmological observations. On this regard, one could also check Refs.\cite{capozziello/2011,capozziello/2008}.

In the paper we have studied one of these alternative theories, the $f(R,T)$ theory. The dependency on $T$ in such a theory is motivated by quantum effects \cite{lobato/2019} and the possible existence of imperfect fluids in the universe. A consequence of the $T$-dependence is the non-conservation of the energy-momentum tensor, which can be evaded from the approaches presented in Section \ref{sec:2}. 

The present literature contained EMC $f(R,T)$ models in cosmology and hydrostatic equilibrium configurations of neutron stars. In both cases, the physical features obtained are significantly more desirable than the non-conservative cases. 

Take, for instance, the EMC $f(R,T)\sim T^{1/2}$ cosmological model, derived in \cite{alvarenga/2013} and observationally tested in \cite{velten/2017}. It has been shown in \cite{velten/2017} that the EMC $f(R,T)$ cosmological model is the only one that passes cosmological tests such as the confrontation with supernovae Ia observational data.

In parallel, in \cite{scmm/2018}, the macroscopical features of neutron stars obtained for the EMC $f(R,T)$ gravity are in touch with massive pulsars observations \cite{demorest/2010,antoniadis/2013}, while the non-conservative case is not \cite{mam/2016}.

Here we have implemented a method to find the conservative  functional form within the $f(R,T)$ function for SSs. We, then, have derived and solved the referred TOV-like equations. 

One may wonder about the dependence of the function $h(T)$ on the EoS and the reliability of such a  feature. This may be due to the geometry-matter coupling predicted by the $f(R,T)$ theory. Note that the inception of material terms in the gravitational action of a given theory yields the possibility of non-minimally couple geometry to matter. In this way, in a fundamental level, the geometrical aspect of the function that shall replace $R$ in the gravitational action is expected to depend on the material features of the system. In this way, different E'soS yield different functional forms for $h(T)$. It is important to quote here that geometry-matter coupling  have shown to yield the cosmic acceleration \cite{zaregonbadi/2016b} and to mimic the dark matter effects \cite{zaregonbadi/2016,bertolami/2010,harko/2010b}.  

Regarding the results obtained for the equilibrium configurations of SSs, in the left panel of Fig.\ref{massxradiusmit} we have seen that for negative values of $\beta$, larger and more massive stars are obtained with the increasing of $|\beta|$. On the other hand, for positive values of $\beta$, we obtain smaller and less massive stars according to the increasing of the parameter.

In the right panel of Fig.\ref{massxradiusmit} we have plotted the star mass against its central energy density for approximately the interval $250-2000$ Mev/fm$^{3}$ of  the latter. Also in Fig.\ref{massxradius} the left panel refers to our EMC model while the right panel is obtained from the $f(R,T)=R+2\chi T$ non-conservative model, as the one of Ref.\cite{das/2016} (although the referred authors have considered a different value for $\omega$), among others \cite{carvalho/2017,mam/2016}. Our conservative model presents a more sensitive contribution to the increasing of mass for the changes in $\lambda<0$ in comparison with the non-conservative case. We also see in Fig.\ref{massxdensity2} that the maximum mass points of our EMC model are attained for smaller values of $\rho_c$ when compared to the results of Ref.\cite{das/2016}. 

Furthermore, the left panel of Fig.\ref{massxdensity2}, together with the regular criterion of stability, indicates a lower limit for $\lambda$ in the present model, which reads $\lambda>-1.5\times 10^{-4}$ and is in agreement with the constraint found in Ref.\cite{carvalho/2017}, being more stringent than the latter by a factor of $2$. 

In what concerns Fig.\ref{massxradiusmit}, which is related to the MIT bag model EoS, with non-null bag constant, we see that for negative values of $\beta$, more massive and greater stars are obtained when the results are confronted to General Relativity. In particular, we have shown the feasibility of our model by comparing our theoretical values of mass and radius with observational data of strange star star candidates and with the results of the non-conserved $f(R,T)$ models as indicated by tables \ref{tab1} and \ref{tab2}.

\acknowledgments
GAC thanks Coordena\c c\~ao de Aperfei\c coamento de Pessoal de N\'ivel Superior (CAPES) grants PDSE 88881.188302/2018-01 and PNPD 88887.368365/2019-00. PHRSM would like to thank S\~ao Paulo Research Foundation (FAPESP), grants
2015/08476-0 and 2018/20689-7, for financial support.




\begin{thebibliography}{99}

\bibitem{weinberg/1989} S. Weinberg, Rev. Mod. Phys. {\bf 61} (1989) 1.

\bibitem{hinshaw/2013} G. Hinshaw et al., Astrophys. J. Suppl. {\bf 208} (2013) 19.

\bibitem{danieli/2019} S. Danieli et al., Astrophys. J. Lett. {\bf 874} (2019) L12.

\bibitem{demorest/2010} P. Demorest et al., Nature {\bf 467} (2010) 1081.

\bibitem{antoniadis/2013} J. Antoniadis et al., Science {\bf 340} (2013) 6131.


\bibitem{howell/2006} D.A. Howell et al., Nature {\bf 443} (2006) 308.

\bibitem{scalzo/2010} R.A. Scalzo et al., Astrophys. J. {\bf 713} (2010) 1073.

\bibitem{kepler/2007} S.O. Kepler et al., Mon. Not. R. Astron. Soc. {\bf 375} (2007) 1315.

\bibitem{nojiri/2011} S. Nojiri and S.D. Odintsov, Phys. Rep. {\bf 505} (2011) 59.

\bibitem{sotiriou/2010} T.P. Sotiriou and V. Faraoni, Rev. Mod. Phys. {\bf 82} (2010) 451.

\bibitem{de_felice/2010} A. De Felice and S. Tsujikawa, Liv. Rev. Rel. {\bf 13} (2010) 161.

\bibitem {riess/1998}A.G. Riess et al., Astron. J. \textbf{116} (1998) 1009.

\bibitem {perlmutter/1999}S. Perlmutter et al., Astrophys. J. \textbf{517}
(1999) 5.

\bibitem{amendola/2007} L. Amendola et al., Phys. Rev. D {\bf 75} (2007) 083504.

\bibitem{song/2007} Y.-S. Song et al., Phys. Rev. D {\bf 76} (2007) 063517.

\bibitem{capozziello/2016} S. Capozziello et al., Phys. Rev. D {\bf 93} (2016) 023501.

\bibitem{astashenok/2013} A.V. Astashenok et al., J. Cosm. Astrop. Phys. {\bf 12} (2013) 040.

\bibitem{astashenok/2015} A.V. Astashenok et al., Phys. Lett. B. {\bf 742} (2015) 160.

\bibitem{das/2015} U. Das and B. Mukhopadhyay, J. Cosm. Astrop. Phys. {\bf 05} (2015) 045.

\bibitem {erickcek/2006} A.L. Erickcek et al., Phys. Rev. D \textbf{74} (2006) 121501.

\bibitem {chiba/2007} T. Chiba et al., Phys. Rev. D \textbf{75} (2007) 124014.

\bibitem {capozziello/2007} S. Capozziello et al., Phys. Rev. D \textbf{76} (2007) 104019.

\bibitem{dolgov/2003} A.D. Dolgov and M. Kawasaki, Phys. Lett. B {\bf 573} (2003) 1.

\bibitem{chiba/2003} T. Chiba, Phys. Lett. B {\bf 575} (2003) 1.

\bibitem{olmo/2005} G.J. Olmo, Phys. Rev. D \textbf{72} (2005) 083505.

\bibitem {harko/2011}T. Harko et al., Phys. Rev. D \textbf{84} (2011) 024020.

\bibitem{xu/2016} M.-X. Xu et al., Eur. Phys. J. C \textbf{76} (2016) 449.

\bibitem{shabani/2014} H. Shabani and M. Farhoudi, Phys. Rev. D {\bf 90} (2014) 044031.

\bibitem{zaregonbadi/2016} R. Zaregonbadi et al., Phys. Rev. D {\bf 94} (2016) 084052.

\bibitem{alhamzawi/2016} A. Alhamzawi and R. Alhamzawi, Int. J. Mod. Phys. D {\bf 25} (2016) 1650020.

\bibitem{baffou/2017} E.H. Baffou et al., Chin. J. Phys. {\bf 55} (2017) 467.

\bibitem {mrc/2016}P.H.R.S. Moraes, G. Ribeiro and R.A.C. Correa, Astrophys.
Space Sci. \textbf{361} (2016) 227.


\bibitem{ms/2017} P.H.R.S. Moraes and P.K. Sahoo, Eur. Phys. J. C {\bf 77} (2017) 480.



\bibitem{barrientos/2014} J. Barrientos O. and G.F. Rubilar, Phys. Rev. D \textbf{90} (2014) 028501.

\bibitem {harko/2014}T. Harko, Phys. Rev. D \textbf{90} (2014) 044067.

\bibitem{harko/2010} T. Harko and F.S.N. Lobo, Eur. Phys. J. C \textbf{70} (2010) 373.

\bibitem{harko/2013} T. Harko et al., Phys. Rev. D \textbf{87} (2013) 047501.



\bibitem{tolman/1939} R.C. Tolman, Phys. Rev. {\bf 55} (1939) 364.

\bibitem{oppenheimer/1939} J.R. Oppenheimer and G.M. Volkoff, Phys. Rev. {\bf 55} (1939) 374.



\bibitem{itoh/1970} N. Itoh, Prog. Theor. Phys. {\bf 44} (1970) 291.

\bibitem{alcock/1986} C. Alcock et al., Astrophys. J. {\bf 310} (1986) 261.

\bibitem{haensel/1986} P. Haensel et al., Astron. Astrophys. {\bf 160} (1986) 121.

\bibitem {li/1999} X.-D. Li et al., Phys. Rev. Lett. \textbf{83} (1999) 3776.

\bibitem {glendenning/1989} L.K. Glendenning, Phys. Rev. Lett. \textbf{63} (1989) 2629.

\bibitem{li/2011} A. Li et al., Res. Astron. Astrophys. {\bf 11} (2011) 482.

\bibitem{bombaci/2002} I. Bombaci, eConf {\bf C010815} (2002) 29.

\bibitem{abbas/2015} G. Abbas et al., Astrophys. Space Sci. {\bf 359} (2015) 17.

\bibitem{panotopoulos/2017} G. Panotopoulos, Gen. Rel. Grav. {\bf 49} (2017) 69.

\bibitem{debabrata/2019} D. Deb et al., Mon. Not. R. Astron. Soc. {\bf 485} (2019) 5652.

\bibitem{deb/2018} D. Deb et al., J. Cosm. Astrop. Phys. {\bf 03} (2018) 044.

\bibitem{biswas/2019} S. Biswas et al., Ann. Phys. {\bf 401} (2019) 1.

\bibitem{mcr/2018} P.H.R.S. Moraes, R.A.C. Correa, G. Ribeiro, Eur. Phys. J. C {\bf 78} (2018) 192.

\bibitem {chavanis/2015} P.-H. Chavanis, Phys. Rev. D \textbf{92} (2015) 103004.


\bibitem{chakraborty/2013} S. Chakraborty, Gen. Rel. Grav. {\bf 45} (2013) 2039.

\bibitem {alvarenga/2013} F.G. Alvarenga et al., Phys. Rev. D \textbf{87} (2013) 103526.

\bibitem{scmm/2018} S.I. dos Santos Jr., G.A. Carvalho, P.H.R.S. Moraes, C.H. Lenzi and M. Malheiro, Eur. Phys. J. Plus {\bf 134} (2019) 398.

\bibitem{tooper/1964} R.F. Tooper, Astrophys. J {\bf 140} (1964) 434.

\bibitem {mam/2016}P.H.R.S. Moraes, J.D.V. Arba\~nil and M. Malheiro, J. Cosmol. Astropart. Phys. \textbf{06} (2016) 005.

\bibitem{velten/2017} H. Velten and T.R.P. Caram\^es, Phys. Rev. D {\bf 95} (2017) 123536.

\bibitem{andersson/2002} N. Andersson et al., Month. Not. Roy. Astron. Soc. {\bf 337} (2002) 1224.

\bibitem{geng/2015} J.J. Geng et al., Astrophys. J {\bf 804} (2015) 21.

\bibitem{mm/2014} P.H.R.S. Moraes and O.D. Miranda, Month. Not. Roy. Astron. Soc. {\bf 445} (2014) L11.

\bibitem{zhou/2018} E.-P. Zhou et al., Phys. Rev. D {\bf 97} (2018) 083015.

\bibitem{malheiro4/2007} F. Weber et al., Int. Jorn. of Mod. Phys. E {\bf 04} (2007) 1165.

\bibitem{malheiro3/2016} J.D.V. Arba\~nil et al., J. Cosm. Astrop. Phys.  {\bf 11} (2016) 012.

\bibitem{malheiro0/2015} J.D.V. Arba\~nil et al., Phys. Rev. D {\bf 92} (2015) 084009.

\bibitem{malheiro2/2003} M. Malheiro et al., J. Phys. G {\bf 29} (2003) 1045.

\bibitem{negreiros/2009} R.P. Negreiros, F. Weber, M. Malheiro and V. Usov,
Phys. Rev. D {\bf 80} (2009) 083006.

\bibitem{carvalho/2018} G.A. Carvalho, R.M. Marinho Jr. and M. Malheiro, Gen. Relativ. Gravit. \textbf{50} (2018) 38.





\bibitem{das/2016} A. Das et al., Eur. Phys. J. C {\bf 76} (2016) 654.

\bibitem{rawls/2011} M.L. Rawls et al., Astrophys. J. {\bf 730} (2011) 25.

\bibitem{guver/2010} T. G\"uver, F. \"Ozel, A. Cabrera-Lavers and P. Wroblewski, Astrophys. J. {\bf 712} (2010) 964.

\bibitem{freire/2011} P.C.C. Freire et al., Mon. Not. R. Astron. Soc. {\bf 412} (2011) 2763.

\bibitem{guver/2010a} T. G\"uver, P. Wroblewski, L. Camarota and F. \"Ozel, Astrophys. J. {\bf 719} (2010) 1807.

\bibitem{ozel/2009} F. \"Ozel, T. G\"uver, and D. Psaltis, Astrophys. J. {\bf 693} (2009) 1775.

\bibitem{elebert/2009} P. Elebert et al., Mon. Not. R. Astron. Soc. {\bf 395} (2009) 884.

\bibitem{abu/2008} M.K. Abubekerov, E.A. Antokhina, A.M. Cherepashchuk and V.V. Shimanskii, Astron. Rep. {\bf 52} (2008) 379.
 

\bibitem{deb/2018b} D. Deb, B.K. Guha, F. Rahaman and Saibal Ray, Phys. Rev. D \textbf{97} (2018) 084026.

\bibitem{capozziello/2011} S. Capozziello and M. de Laurentis, Phys. Rep. {\bf 509} (2011) 167.

\bibitem{capozziello/2008} S. Capozziello and M. Francaviglia, Gen. Rel. Grav. {\bf 40} (2008) 357.

\bibitem{lobato/2019} R.V. Lobato, G.A. Carvalho, A.G. Martins and P.H.R.S. Moraes, Eur. Phys. J. Plus \textbf{134}, (2019) 132.

\bibitem{zaregonbadi/2016b} R. Zaregonbadi and M. Farhoudi, Gen. Relativ. Gravit. \textbf{48} (2016) 142.

\bibitem{bertolami/2010} O. Bertolami and J. P\`aramos, J. Cosm. Astrop. Phys. {\bf 03} (2010) 009.

\bibitem{harko/2010b} T. Harko, Phys. Rev. D {\bf 81} (2010) 084050.

\bibitem{carvalho/2017} G.A. Carvalho, R.V. Lobato, P.H.R.S. Moraes, J.D.V. Arba\~nil, E. Otoniel, R. Marinho Jr. and M. Malheiro , Eur. Phys. J. C {\bf 77} (2017) 871.













\end{thebibliography}
\end{document}